\documentclass[prd,twocolumn,superscriptaddress,altaffilletter,showpacs,nofootinbib]{revtex4}
\usepackage[dvips]{graphicx}
\usepackage{amsmath}

\newcommand{\be}{\begin{equation}}
\newcommand{\ee}{\end{equation}}
\newcommand{\bea}{\begin{eqnarray}}
\newcommand{\eea}{\end{eqnarray}}

\begin{document}

\title{Randall-Sundrum brane cosmology: modification of late-time cosmic dynamics by exotic matter}

\author{Ricardo Garc\'{\i}a-Salcedo}\email{rigarcias@ipn.mx}\affiliation{Centro de Investigacion en Ciencia Aplicada y Tecnologia Avanzada - Legaria
del IPN, M\'{e}xico D.F., M\'{e}xico.}

\author{Tame Gonzalez}\email{tamegc72@gmail.com}\affiliation{Departamento de F\'{\i}sica, Centro de Investigacion y de Estudios Avanzadas del IPN, A.P.14-740,07000, M\'exico D.F., M\'exico.}

\author{Claudia Moreno}\email{claudia.moreno@cucei.udg.mx}\affiliation{Departamento de F\'{\i}sica y Matem\'aticas, Centro Universitario de
Ciencias Ex\'actas e Ingenier\'{\i}as, Corregidora 500 S.R., Universidad de Guadalajara, 44420 Guadalajara, Jalisco, M\'exico}

\author{Israel Quiros}\email{iquiros@fisica.ugto.mx}\affiliation{Divisi\'on de Ciencias e Ingenier\'ia de la Universidad de Guanajuato, A.P. 150, 37150, Le\'on, Guanajuato, M\'exico.}

\date{\today}

\begin{abstract}
In this paper we show, through the study of concrete examples, that, depending on the cosmic dynamics of the energy density of matter degrees of freedom living in the brane, Randall-Sundrum (RS) brane effects can be important not only at short distances (UV regime), but also at large cosmological scales (IR regime). Our first example relies on the study, by means of the dynamical systems tools, of a toy model based in a non-linear electrodynamics (NLED) Lagrangian. Then we show that other, less elaborated models, such as the inclusion of a scalar phantom field, and of a tachyon phantom field -- trapped in the brane -- produce similar results. The form of the RS correction seems to convert what would have been future attractors in general relativity into saddle points. The above "mixing of scales" effect is distinctive only of theories that modify the right-hand-side (matter part) of the Friedmann equation, so that, for instance, DGP-brane models do not show this feature.
\end{abstract}

\maketitle

\section{Introduction}

Braneworld models of the Universe, where standard model (SM) particles are trapped in a three-space, while gravity might propagate also in the extra-space, have been the focus of attention in the current decade in the search for answers to several outstanding problems in theoretical physics and in cosmology. The Dvali-Gabadadze-Porrati (DGP) brane model \cite{dgp}, for instance, was explored in an attempt to understand the mystery underlaying the present accelerating pace of the cosmic expansion, without invoking the mysterious "dark energy". The DGP brane model modifies general relativity (GR) at large cosmological scales or, in other words, it carries infra-red (IR) modifications of GR. In this model the present speedup of the cosmic expansion is explained as an effect of the extra-dimension: at large cosmological scales gravity leaks into the extra space, resulting in a weakening of the gravitational bounds of the cosmic structure.\footnote{The DGP models have a ghost mode in the self-accelerating branch which raises serious doubts about it being a competing alternative to the dark energy \cite{k-koyama}.} Acting on the opposite end of the energy/curvature spectrum is the Randall-Sundrum (RS) model \cite{rs1,rs2}, which carries ultra-violet (UV) modifications of GR. This time, however, the motivation was to seek for an alternative explanation to the hierarchy problem (RS1 model \cite{rs1}), and to look for an alternative to the Kaluza-Klein compactification procedure (RS2 model \cite{rs2}). The model known as RS2 found important applications in cosmology \cite{deffayet}, in particular for the inflationary scenario \cite{hawkin-lindsay}. In this model a single codimension-1 brane with positive tension, is embedded in a five-dimensional anti-de Sitter (AdS) bulk spacetime, which is infinite in the direction perpendicular to the brane. As said, the standard model matter degrees of freedom are confined to the brane, while gravitation can propagate in the bulk. However, in the low-energy/IR limit, due to the curvature of the bulk, the graviton is confined to the brane, and standard (four-dimensional) GR laws are recovered. The latter result can be better understood if we invoke a cosmological application \cite{deffayet}: let us consider a homogeneous and isotropic Friedmann-Robertson-Walker (FRW) metric with flat spatial sections. The Friedmann equation on the RS2 brane then reads (we neglect the ''dark radiation'' and cosmological constant terms): $$3H^2=\rho(1+\rho/2\lambda),$$ where $\rho$ is the energy density of the matter degrees of freedom trapped in the brane, and $\lambda$ is the brane tension. The latter is fixed by the curvature scale of the bulk, which is constrained to be less than 0.1 mm by current tests of Newton's law \cite{roy-kazuya}. At very high energy density ($\rho\gg\lambda$), due to the brane effects, the above equation is fundamentally modified: $H\propto\rho$. As the cosmic expansion proceeds the energy content in the brane dilutes and, eventually, as long as $\rho\ll\lambda$, the standard GR behavior $H^2\propto\rho$ is recovered. 

It is quite obvious that the latter result is highly dependent on the cosmic dynamics of the energy density $\rho$ itself. Actually, for fields whose energy density increases in the course of the cosmic expansion, the brane effects can be important at late-times/large cosmological scales also. This simple and obvious fact -- when analyzed in the light of the cosmological dynamics of fields trapped in the brane -- seems to contradict the spirit of the Randall-Sundrum brane scenario, which is usually assumed to modify gravity at high UV energy/curvature, with special impact on primordial (early) inflation scenario. Our trivial finding implies that, if the energy density of some (perhaps exotic) component of the cosmic mixture does not dilute but grows as expansion proceeds, then the corresponding RS-brane model can have implications not only for early-time cosmology, but it can have appreciable impact also on the fate of the cosmic expansion.

That is why, in the present paper we will explore the impact of RS brane effects on the cosmic dynamics, when the energy density of -- at least one of the components of -- the matter degrees of freedom trapped in the brane grows with the course of the cosmological expansion, instead of diluting itself. To this end we will explore the asymptotic properties of concrete models of very different nature, starting with a very elaborated toy model based on a non-linear electrodynamics (NLED) Lagrangian whose asymptotic structure is very rich. By investigating how the brane effects modify the asymptotic structure of the latter model, we will show that RS2 brane effects might convert what would have been future attractors in GR into saddle critical points. This result will be corroborated through the study of other examples of exotic matter trapped in the RS brane: i) a scalar phantom field, and ii) a tachyon phantom field. The physical consequences of the above curious ("trivial") finding will be discussed.

\section{NLED-based toy model}

As already mentioned, we start by exploring the dynamics of a non-linear electrodynamics (NLED)-based toy model. The four-dimensional (4D) Einstein-Hilbert action of gravity coupled to NLED is given by: 

\be S=\int d^4x\sqrt{-g}\left[R+L_\gamma+L(F,G)\right],\label{nled action}\ee where $R$ is the curvature scalar, $L_\gamma$ -- the background perfect fluid's Lagrangian, and $L(F,G)$ is the gauge-invariant electromagnetic Lagrangian, which is a function of the electromagnetic invariants $F\equiv F^{\mu \nu}F_{\mu \nu }$ and $G\equiv\frac{1}{2}\epsilon_{\alpha\beta\mu\nu}F^{\alpha\beta}F_{\mu\nu}$ \cite{novellocqg}. As usual, the electromagnetic tensor is defined as: $F_{\mu\nu}=\partial_\mu A_\nu-\partial_\nu A_\mu$. Standard (linear) Maxwell electrodynamics is given by the Lagrangian $L(F)=-F/4$. The corresponding field equations can be derived from the action by performing variations with respect to the spacetime metric $g_{\mu\nu}$, to obtain: 

\be G_{\mu\nu}=T_{\mu\nu}^\gamma+T_{\mu\nu}^{EM},\label{feqs}\ee where 

\bea &&T_{\mu\nu}^\gamma=(\rho_\gamma+p_\gamma) u_\mu u_\nu-p_\gamma g_{\mu\nu},\nonumber\\&&T_{\mu\nu}^{EM}=g_{\mu\nu}[L(F)-G L_G]-4F_{\mu\alpha}F_\nu^{\;\;\alpha} L_F,\nonumber\eea with $\rho_{\gamma}=\rho_{\gamma}(t)$, $p_\gamma=p_\gamma(t)$ -- the energy density and barotropic pressure of the background fluid, respectively, while $L_F\equiv dL/dF$, $L_{FF}\equiv d^2L/dF^2$, etc. Variation with respect to the components of the electromagnetic potential $A_\mu$ yields to the electromagnetic field equations. 

In order to meet the requirements of homogeneous and isotropic cosmologies (as, in particular, the one associated with FRW spacetimes), the energy density and the pressure of the NLED field should be evaluated by averaging over volume. To do this, we define the volumetric spatial average of a quantity $X$ at the time $t$ by  (for details see \cite{novellocqg} and references therein): $$\overline{X}\equiv \lim_{V\rightarrow V_{0}}\frac{1}{V}\int d^3x \sqrt{-g}\;X\;,$$ where $V=\int d^3x\sqrt{-g}$ and $V_{0}$ is a sufficiently large time-dependent three-volume. Following the above averaging procedure, for the electromagnetic field to act as a source for the FRW model we need to impose that (the Latin indexes run over three-space); $\overline{E}_{i}=0$, $\overline{B}_{i}=0$, $\overline{E_{i}B_{j}}=0$, and also $3\overline{E_i E_j}=-E^2 g_{ij}$, $3\overline{B_i B_j}=-B^{2}g_{ij}$. Additionally it has to be assumed that the electric and magnetic fields, being random fields, have coherent lengths that are much shorter than the cosmological horizon scales. Under these assumptions the energy-momentum tensor of the electromagnetic (EM) field -- associated with the Lagrangian density $L=L(F,G)$ --, can be written in the form of the energy-momentum tensor for a perfect fluid: $$T_{\mu\nu}^{EM}=\left(\rho_B+p_B\right) u_{\mu}u_{\nu}-p_Bg_{\mu\nu},$$ where $\rho_B=-L+G L_G-4L_F E^2$, and $p_B=L-G L_G-4(2B^2-E^2) L_F/3$, with $E$ and $B$ being the averaged electric and magnetic fields, respectively. In what follows, to simplify the analysis, we shall consider a FRW universe filled with a ''magnetic fluid'', i. e., the electric component $E$ will be assumed vanishing. Even this simplified picture can give important physical insights. We will be focusing in the study of the asymptotic properties of a cosmological model with interesting features, namely a phase of current cosmic acceleration and the absence of an initial singularity, which was proposed in \cite{novellocqg} (see also \cite{23}) and is based upon the following Lagrangian density:

\be L=-\frac{1}{4}F+\alpha F^2+\beta F^{-1},  \label{L}\ee where $\alpha$ and $\beta$ are arbitrary (constant) parameters. As seen this Lagranagian contains both positive and negative powers of $F$. The second (quadratic) term dominates during very early epochs of the cosmic dynamics, while the Maxwell term (first term above) dominates in the radiation era. The last term in (\ref{L}) is responsible for the accelerated phase of the cosmic evolution \cite{novellocqg}. The above Lagrangian density yields a unified scenario to describe both the acceleration of the universe (for weak fields) and the avoidance of the initial singularity, as a consequence of its properties in the strong-field regime. Recalling that we are considering magnetic universes, i.e., $F=2B^2$, where $B^2$ is an averaged value of the magnetic field,\footnote{For details of the averaging procedure consult \cite{novellocqg}.} the stress-energy tensor associated with (\ref{L}) can be written in the form of an equivalent perfect fluid stress-energy tensor with energy density and parametric pressure: $\rho_B=B^2(1-8\alpha B^2-\beta B^{-4})/2$, and $p_B=B^2(1-40\alpha B^2+7\beta B^{-4})/6$, respectively.\footnote{Notice that, for large values of the NLED field, positivity of energy requires that $B<1/\sqrt{8\alpha }$, while, for small enough values of $B\ll 1$, if one considers positive $\beta>0$, then positivity of energy implies that $B>(7\beta)^{1/4}$. The existence of the lower bound, at first sight might appear problematic, however, given that the observational data constraints the parameter $\beta$ to be $\sqrt{|\beta|}\approx 4\times 10^{-28}\;g\;cm^{-3}$ \cite{23}, then the lower bound on $B$ can be admitted without going into conflicts with observations.} 

In this paper we assume the background fluid to be dust cold dark matter (CDM), so that $\gamma=1$. Our goal will be to put the corresponding cosmological equations: 

\begin{eqnarray} &&3H^2=\rho_{cdm}+\rho_B,\;\;2\dot H=-\rho_{cdm}-(\rho_B+p_B),\nonumber\\
&&\dot\rho_{cdm}+3H\rho_{cdm}=0,\;\;\dot\rho_B+3H(\rho_B+p_B)=0,\label{nled feqs}\end{eqnarray} in the form of an autonomous system of ordinary differential equations (ODE). For this purpose we choose the following phase space variables: $$x\equiv\frac{\rho_B}{3H^2},\;\;y\equiv\frac{16\alpha B^4}{3H^2},\;\;z\equiv\frac{4\beta}{3H^2 B^2}\;.$$ After this choice, the following autonomous system of ODE is obtained out of (\ref{nled feqs}):

\bea &&x'=(x-1)(x-y+z),\;\;y'=-y(5-x+y-z),\nonumber\\
&&z'=z(7+x-y+z).\label{asode}\eea Here, for generality of the analysis we shall consider arbitrary $\alpha\in\Re$ and $\beta\in\Re$. In what follows in this paper the tilde will denote derivative with respect to the new time-variable $\tau\equiv\ln a\;(d\tau=H dt)$, also known as the number of e-foldings. The phase space relevant to the present study is then given by the following region in $(x,y,z)$: $$\Psi_U=\{(x,y,z)|0\leq x\leq 1,\;(y,z)\in\Re^2,\;8x+2y+z\geq 0\},$$ where we have considered the fact that $B^2/6H^2=x+y/4+z/8$. In what follows we use the standard definition of the dimensionless density parameter for the component $X$: $\Omega_X\equiv\rho_X/3H^2$, while $\omega_X=p_X/\rho_X$ will denote the effective equation of state (EOS) parameter of the component $X$. Four equilibrium points of (\ref{asode}) in $\Psi_U$, are found:

\begin{enumerate}

\item Radiation-dominated phase: $$P_{rad}=(x,y,z)=(1,0,0),\;\;\Omega_B=1,\;\;\Omega_{cdm}=0\;.$$ This is a decelerating expansion solution ($q=1$), that is fueled by standard radiation with $\omega_B=1/3$. The eigenvalues of the linearization matrix corresponding to this equilibrium point are: $\lambda_1=1$, $\lambda_2=-4$, $\lambda_3=8$, so that it is a saddle in $\Psi_U$.

\item Infra-red NLED-dominated solution: $$P_{nled}^{IR}=(1,0,-8),\;\;\Omega_B=1,\;\;\Omega_{cdm}=0\;.$$ This is a late-time, super-inflationary solution ($q=-3$), where the NLED fluid mimics phantom behavior ($\omega_B=-7/3$). This solution exist only for $z<0$ (negative $\beta<0$). In this case there is no lower bound on the magnitude of the magnetic field. The eigenvalues of the linearization matrix for $P_{nled}^{IR}$ are: $\lambda_1=-8$, $\lambda_2=-7$, $\lambda_3=-12$, so that this solution corresponds to a late-time (future) attractor. Since, the NLED-magnetic field mimics phantom behavior, the late-time attractor might be associated with a cosmological singularity (most probably a big-rip type of singularity).

\item Ultra-violet NLED-dominated phase: $$P_{nled}^{UV}=(1,-4,0),\;\;\Omega_B=1,\;\;\Omega_{cdm}=0\;.$$ This solution corresponds to an early-time, super-stiff-fluid solution ($\omega_B=5/3$), which is associated with super-decelerating expansion ($q=3$). This solution exist only for $y<0$, i. e., if the constant $\alpha$ is a negative quantity ($\alpha<0$). Curiously, this case does not meet the conditions for a bounce (there is no upper bound on the magnitude of the magnetic field), and the corresponding cosmology starts with a big-bang singularity, since $B^2/H^2=0,\;B\neq 0,\;\Rightarrow\;H\rightarrow\infty$. This solution is a past attractor in $\Psi_U$, since the eigenvalues of the corresponding linearization matrix: $\lambda_1=5$, $\lambda_2=12$, $\lambda_3=4$, are all positive quantities. This means that this point is the starting point of every probe path in the phase space $\Psi_U$.

\item CDM-dominated solution: $$P_{cdm}=(0,0,0),\;\;\Omega_B=0,\;\;\Omega_{cdm}=1\;.$$ This phase of the cosmic evolution is characterized by decelerated expansion ($q=1/2$). The EOS parameter for the magnetic field is undefined in this case. The existence of this solution is necessary for the formation of the observed amount of structure. This is also a saddle equilibrium point in $\Psi_U$, since: $\lambda_1=-5$, $\lambda_2=7$, $\lambda_3=-1$.

\end{enumerate}

For positive definite $\alpha$ and $\beta$, only the radiation-dominated equilibrium point $P_{nled}$, and the CDM-dominated solution $P_{cdm}$, are found in $\Psi_U$. The above results confirm our expectation that a combination of positive and negative powers of the electromagnetic invariant $F$ in the NLED Lagrangian, can drive a very interesting cosmological scenario leading to accelerated expansion at late times. Unfortunately, early-time (primordial) inflation can not be obtained in the present model.

\subsection{NLED on the RS brane}

Now we explore the possible effect of RS2 braneworld gravity on the above picture. The corresponding (modified) FRW cosmological equations for a RS2 brane with CDM and a perfect fluid of NLED trapped in it, can be written as (here we omit the ''dark radiation'' term);

\bea &&3H^2=\rho_{tot}\left(1+\frac{\rho_{tot}}{2\lambda}\right),\;\;\rho_{tot}=\rho_{cdm}+\rho_B,\nonumber\\
&&2\dot H=-(\rho_{cdm}+\rho_B+p_B)\left(1+\frac{\rho_{tot}}{\lambda}\right),\nonumber\\
&&\dot\rho_{cdm}=-3H\rho_{cdm},\;\;\dot B=-2H B.\label{feqs''}\eea It is convenient to introduce the following phase space variables: $$x\equiv\frac{\rho_B}{3H^2},\;\;y\equiv\frac{16\alpha B^4}{3H^2},\;\;z\equiv\frac{4\beta}{3H^2 B^2},\;\;v\equiv\frac{\rho_{tot}}{3H^2}\;.$$ The variable $v$ controls the brane regime, 

$$\Omega_\lambda\equiv\frac{\lambda}{3H^2}=\frac{v^2}{2(1-v)},$$ so that, for instance, the GR-limit (formal limit $\lambda\rightarrow\infty$) corresponds to $v=1$. The following autonomous system of ODE can be derived out of (\ref{feqs''}):

\bea &&x'=3(1-v)x+\left[\left(\frac{2-v}{v}\right)x-1\right](x-y+z),\nonumber\\
&&y'=y\left[-5+(1-v)+\left(\frac{2-v}{v}\right)(x-y+z)\right],\nonumber\\
&&z'=z\left[7+(1-v)+\left(\frac{2-v}{v}\right)(x-y+z)\right],\nonumber\\
&&v'=(1-v)(3v+x-y+z).\label{brane asode}\eea At $v=1$ the first three equations above coincide with the equations (\ref{asode}), which hold for GR with a NLED-magnetic field. The phase space of the model can be defined as follows: 

\bea &&\Psi_U^{brane}=\{(x,y,z,v)|\;0\leq x\leq 1,\;\;(y,z)\in\Re^2,\nonumber\\&&\;\;\;\;\;\;\;\;\;\;\;\;\;\;\;\;\;\;\;\;\;\;\;\;\;\;\;\;8x+2y+z\geq 0,\;0\leq v\leq 1\}.\nonumber\eea 

The critical points of the autonomous system of ODE (\ref{brane asode}) in the phase space $\Psi_U^{brane}$, their physical properties and stability are discussed below. As in the general relativity case, four equilibrium points are found:

\begin{enumerate}

\item CDM-dominated solution: $$P_{cdm}=(0,0,0,1),\;\;\Omega_B=0,\;\;\Omega_{cdm}=1\;.$$ Since $q=1/2$, this solution is associated with decelerated expansion. The NLED-magnetic field EOS parameter $\omega_B$ is undefined. This is a saddle equilibrium point in $\Psi_U^{brane}$. Actually, the eigenvalues of the linearization matrix corresponding to this point are: $\lambda_1=-5$, $\lambda_2=7$, $\lambda_3=-3$, $\lambda_4=-1$.

\item Radiation-dominated solution: $$P_{rad}=(1,0,0,1),\;\;\Omega_B=1,\;\;\Omega_{cdm}=0\;.$$ It is also a decelerated-expansion solution ($q=1$) driven by standard (Maxwell) radiation ($\omega_B=1/3$). As the CDM-dominated phase, this solution also represents a saddle equilibrium point in $\Psi_U^{brane}$, since the eigenvalues of the linearization matrix are of opposite signs: $\lambda_1=8$, $\lambda_2=1$, $\lambda_{3,4}=-4$.

\item UV NLED-dominated solution: $$P_{nled}^{UV}=(1,-4,0,1),\;\;\Omega_B=1,\;\;\Omega_{cdm}=0\;.$$ This solution shares many properties with its similar GR-solution: it is a super-decelerated ($q=3$), super-stiff state ($\omega_B=5/3$), associated with a big-bang-type singularity ($H\rightarrow\infty$). The new feature here is that the stability properties have been modified by the brane effects. Actually, the eigenvalues of the linearization matrix in the present case are: $\lambda_1=12$, $\lambda_2=4$, $\lambda_3=5$, $\lambda_4=-8$, so that this is a saddle equilibrium point in $\Psi_U^{brane}$ (its similar GR-solution is a past attractor).

\item IR NLED-dominated solution: $$P_{nled}^{IR}=(1,0,-8,1),\;\;\Omega_B=1,\;\;\Omega_{cdm}=0\;.$$ As in the former case, this solution shares many properties with its GR-similar: it is a super-accelerated ($q=-3$), phantom-like solution ($\omega_B=-7/3$), possibly associated with a big-rip-type singularity. Other properties, the stability in particular, have been modified by the brane effects. Actually, the eigenvalues of the Jacobian matrix are $\lambda_1=-12$, $\lambda_2=-8$, $\lambda_3=-7$, $\lambda_4=4$, so that it is also a saddle equilibrium point in the phase space $\Psi_U^{brane}$.

\end{enumerate}

The first thing that is worthy of mention, is the fact that all of the critical points found represent saddle equilibrium points in the phase space $\Psi_U^{brane}$. Note that, since $v=1$, the four equilibrium points are associated with GR. In fact these coincide in almost all aspects with the ones found previously for the GR case. The first two points $P_{cdm}$, and $P_{rad}$, show no fundamental differences with their GR-similar. However, the NLED-dominated solutions $P_{nled}^{UV}$ and $P_{nled}^{IR}$, have different stability properties than their corresponding GR-solutions: while in the latter case $P_{nled}^{UV}$ was the past attractor and $P_{nled}^{IR}$ was the future attractor, in the present case both are saddle equilibrium points as already said. This means, in turn, that the space-time singularities associated with these critical points (big-bang and big-rip singularities respectively), might be evaded in the present case. Modification of the stability properties of the UV solution was expected since, as already mentioned, RS brane effects are appreciable at high energies/short distances (early times). Actually, only at very high energies can the graviton acquire large momenta along the extra dimension and may escape into the bulk (5D) spacetime. The surprise was the IR solution: it is expected that at low energies (large cosmological scales) the RS2 brane effects can be safely ignored. However, while making such statements one has to be careful. In the cosmological context, the most appreciable RS2 brane effect is to modify the Friedmann equation: $3H^2=\rho_{tot}(1+\rho_{tot}/2\lambda)$. Hence, at very high energy density $\rho_{tot}\gg\lambda$ (much bigger than the brane tension), the Friedmann equation is fundamentally modified $3H^2\propto\rho_{tot}^2$. If in the course of the cosmic expansion the total energy content of the universe dilutes, then as long as $\rho_{tot}\ll\lambda$, one recovers standard GR-Friedmann behavior. Now look at the Lagrangian density for the NLED-magnetic field (\ref{L}) considered here. Note that as the expansion proceeds the magnetic field $F\propto B^2$ dilutes, and the component $\propto\beta B^{-2}$ in (\ref{L}) grows without limit. This means that the total energy content of the universe starts growing at the expense of the NLED component so that, at late times, eventually, $\rho_{tot}$ might become much larger than the brane tension once again, rendering the brane effects important at late times also. It is then to be expected that similar IR modifications will be produced if one considers other less elaborated models such as, for instance, a scalar phantom field trapped in the RS2 braneworld.

\begin{table*}[tbp]\caption[crit]{Critical points of the autonomous system of ODE (\ref{T-ode}) for the potential $V(\phi)=V_0\phi^{-2}$ and their properties. These critical points always exist. The parameters $\lambda_1$ and $\lambda_2$ are the eigenvalues of the linearization matrices corresponding to each one of the critical points. We have defined: $y_*=\sqrt{(\lambda^2+\sqrt{\lambda^4+36})/6}$.}
\begin{tabular}{@{\hspace{14pt}}c@{\hspace{14pt}}c@{\hspace{14pt}}c@{\hspace{14pt}}c@{\hspace{14pt}}c@{\hspace{14pt}}c@{\hspace{14pt}}c@{\hspace{14pt}}c@{\hspace{14pt}}c}
\hline\hline\\[-0.3cm]
C. Point &$x$&$y$&$\Omega_{cdm}$ &$\Omega_\phi$& $\omega_\phi$ & $q$ & $\lambda_1$&$\lambda_2$\\[0.1cm]
\hline\\[-0.2cm]
$M$ & $0$ & $0$ & $1$ & $0$ & $-1$ & $1/2$  &$-3$&$3/2$\\[0.2cm]
$TP$&$-\lambda y_*/\sqrt{3}$&$y_*$&$0$ &$1$&$-1-\lambda^2y_*^2/3$&$-1-\lambda^2y_*^2/2$ &$-3-\lambda^2y_*^2/2$&$-3-\lambda^2y_*^2$\\[0.2cm]
\hline \hline
\end{tabular}\label{tab1}
\end{table*}

\section{The case of scalar/tachyon phantom fields}

Consider now that the source of the Einstein's field equations are a background of dark matter and a scalar phantom field $\phi$. In this case the cosmological equations coincide with (\ref{nled feqs}) but with the replacement of $\rho_B$, $p_B$ by: 

\be \rho_\phi=-\frac{\dot\phi^2}{2}+V(\phi),\;\;p_\phi=-\frac{\dot\phi^2}{2}-V(\phi),\label{phantom-rho-p}\ee where $V(\phi)$ is the self-interaction potential. To write the corresponding cosmological equations in the form of an autonomous system of ODE, it suffices to choose the following phase space variables: 

\be x\equiv\frac{\dot\phi}{\sqrt 6 H},\;\;y\equiv\frac{\sqrt V}{\sqrt 3 H}.\label{phantom-var}\ee The following autonomous system of ODE is obtained:

\bea &&x'=\sqrt{\frac{3}{2}}\frac{\partial_\phi V}{V} y^{2}-3x-x\frac{H'}{H},\nonumber\\
&&y'=\sqrt{\frac{3}{2}}\frac{\partial_\phi V}{V} xy-y\frac{H'}{H},\label{phantom-ode}\eea where $H'/H=-3(1-x^2-y^2)/2$. As before we can write the relevant parameters in terms of the variables $x$, $y$:

\be \Omega_\phi=y^2-x^2,\;\;\Omega_{cdm}=1+x^2-y^2,\;\;\omega_\phi=\frac{x^2+y^2}{x^2-y^2}.\nonumber\ee Notice, also, that for the deceleration parameter one has $q=-(1+H'/H)$. The phase space relevant for this case is defined by the following set: $$\Psi_{ph}=\left\{(x,y)|\;0\leq y^2-x^2\leq 1,\;\;y\geq 0\right\}.$$ For simplicity we shall consider an exponential self-interaction potential for the phantom field: $V(\phi)=V_0 \exp(-\lambda\phi)$, where $\lambda$ is a constant. For this potential the system of ODE (\ref{phantom-ode}) is closed. Only two equilibrium points are found in this case:

\begin{enumerate}

\item Matter-dominated point: $$M=(0,0),\;\;\Omega_{cdm}=1,\;\;\Omega_\phi=0,\;\;q=\frac{1}{2},$$ while the equation of state $\omega_\phi$ is undefined. The eigenvalues of the corresponding linearization (Jacobian) matrix are: $\lambda_1=-3/2$, $\lambda_2=3/2$, so that this a saddle equilibrium point.

\item Scalar phantom-dominated solution: 

\bea &&SP=(-\lambda/\sqrt{6},\sqrt{1+\lambda^2/6}),\;\;\Omega_{cdm}=0,\nonumber\\
&&\Omega_\phi=1,\;\;\omega_\phi=-1-\frac{\lambda^2}{3},\;\;q=-1-\frac{\lambda^2}{2}.\nonumber\eea Notice that $\omega_\phi<-1$, $q<-1$ so that this is a supper-accelerated, phantom-dominated solution, independent of whether $\lambda$ is a positive or a negative quantity. This critical point is a future attractor in the phase space $\Psi_{ph}$ since both eigenvalues of the Jacobian matrix are negative in this case: $\lambda_1=-3-\lambda^2/2$, $\lambda_2=-3-\lambda^2$.

\end{enumerate}

Lets turn our attention now to the phantom tachyon field, for which, instead of the definition (\ref{phantom-rho-p}) for the effective energy density and pressure, one has:

\be \rho_\phi=\frac{V}{\sqrt{1+\dot{\phi}^2}}\;,\;\;p_\phi=-V\sqrt{1+\dot{\phi}^2}.\label{T-rho-p}\ee Notice that the sign of the kinetic energy under the square root has been changed. The Klein-Gordon equation for the phantom tachyon is

\be \ddot{\phi}+3H\dot{\phi}\left[1+\dot{\phi}^2\right]-\frac{\partial_\phi V}{V}\left[1+\dot{\phi}^2\right]=0.\label{klein-gordon}\ee In order to write the cosmological equations in the form of an autonomous system of ODE, in this case it is recommendable to work with the following phase space variables \cite{copeland}: 

\be x\equiv\dot{\phi},\;\;y\equiv\frac{\sqrt{V}}{\sqrt{3}H}.\label{T-var}\ee Notice that, while the phase space variables (\ref{phantom-var}) were normalized by the Hubble parameter $H$, the ones in (\ref{T-var}) are not normalized (at least $x$ in (\ref{T-var}) is not a normalized variable). After the above choice, one can obtain the following autonomous system of ODE:

\bea &&x'=(x^2+1)\left[-3x+\sqrt{3}y\;\frac{\partial_\phi V}{V^{3/2}}\right]\nonumber\\
&&y'=y\left[\sqrt{\frac{3}{4}}xy\;\frac{\partial_\phi V}{V^{3/2}}-\frac{H'}{H}\right],\label{T-ode}\eea where $$\frac{H'}{H}=-\frac{3}{2}(1-y^2\sqrt{x^2+1}).$$ In terms of the variables $x$, $y$ in (\ref{T-var}), one has: $\Omega_\phi=y^2/\sqrt{1+x^2}$, while for the phantom tachyon EOS parameter, $\omega_\phi=-1-x^2$. The phase space spanned by the variables $x$ and $y$, that is relevant for an expanding Universe, can be defined as: $$\Psi_T=\{(x,y):-1\leq x\leq 1,\;0\leq y^4\leq 1-x^2\}.$$ In general the system (\ref{T-ode}) is not a closed system of ODE, unless one considers the specific case when $\lambda=\partial_\phi V/V^{3/2}$ is a constant \cite{copeland}, so that $V(\phi)=V_0 \phi^{-2}$ ($V_0=4/\lambda^2$). In what follows, for simplicity, we shall consider, precisely, the latter kind of self-interaction potential for the phantom tachyon. The critical points of (\ref{T-ode}) in this case, as well as their relevant properties, are shown in Tab. \ref{tab1}. As in the former case there are found only two critical points: 

\begin{enumerate}

\item Matter-dominated solution $M=(0,0)$. This is a decelerated expansion-solution and represents a saddle point in the phase space $\Psi_T$.

\item Tachyon phantom-dominated solution $TP=(-\lambda y_*/\sqrt 3,y_*)$. In this case, since both $\omega_\phi<-1$, and $q<-1$, it is a super-accelerated, phantom solution. This equilibrium point is a future attractor in $\Psi_T$.

\end{enumerate}

\begin{figure}[t!]
\begin{center}
\includegraphics[width=6.5cm,height=6cm]{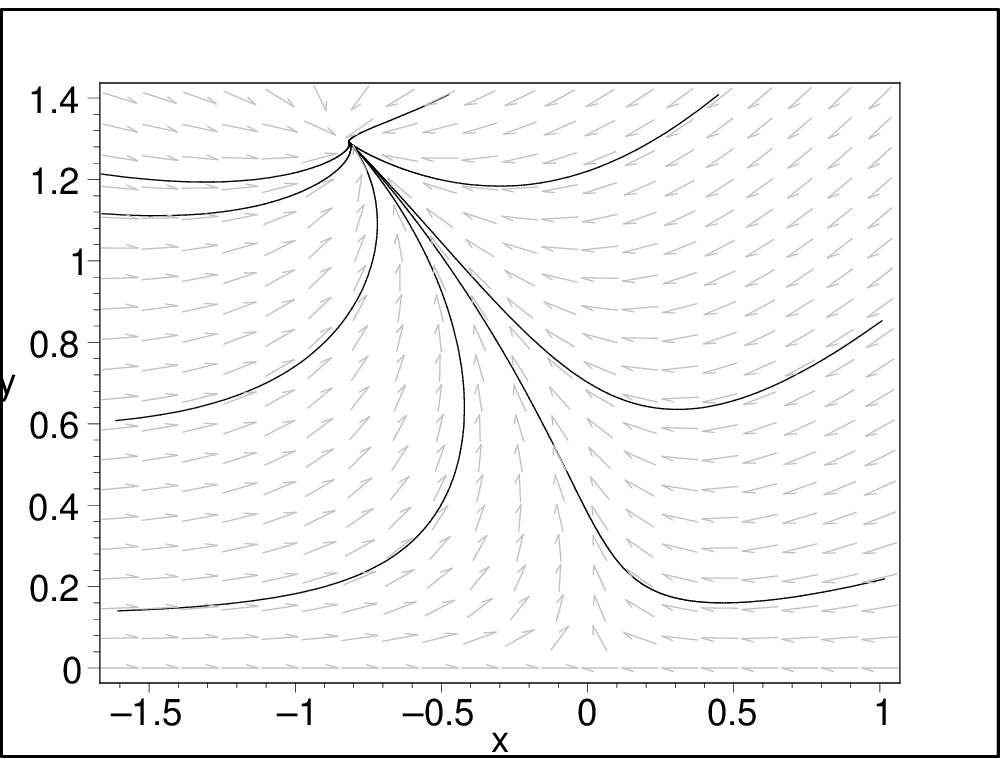}\vspace{0.3cm}
\includegraphics[width=6.5cm,height=6cm]{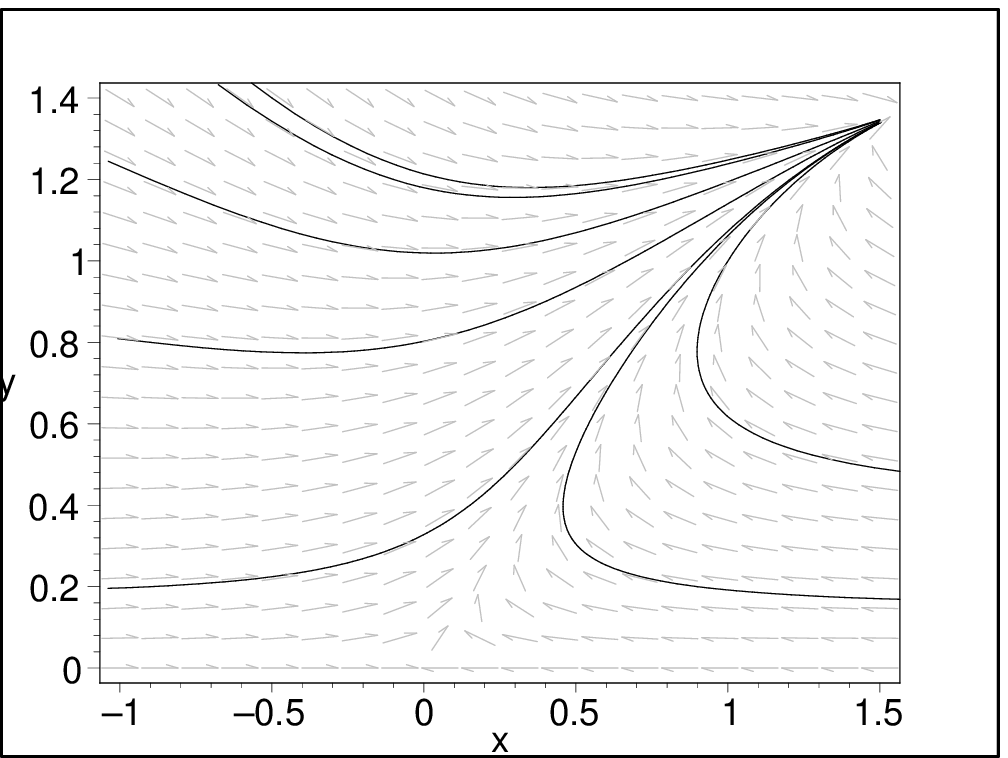}\vspace{0.3cm}
\end{center}
\caption{Phase portraits for the scalar phantom field (left-hand panel), and for the tachyon phantom field (right-hand panel), respectively. The brane tension $\lambda$ has been arbitrarily set equal to $\lambda=2$. In both cases there are no past attractors, while the phantom-dominated solution is always the future attractor.} \label{fig1}
\end{figure}

The above discussed asymptotic structure is clearly illustrated in the Fig. \ref{fig1}, where the phase portraits for the scalar phantom field (left-hand panel), and for the tachyon phantom field (right-hand panel), respectively, are drawn for a given value of the brane tension ($\lambda=2$). It is apparent how the phase trajectories approach to the future attractor: the phantom-dominated solution.

\begin{figure}[t!]
\begin{center}
\includegraphics[width=6.5cm,height=6cm]{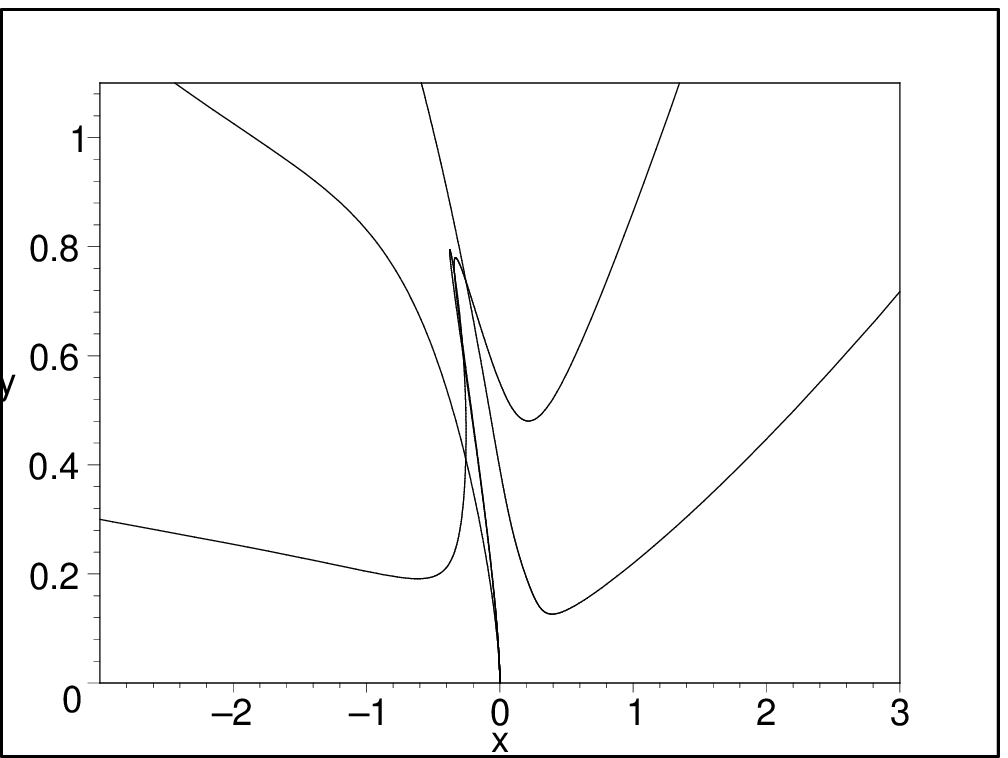}
\includegraphics[width=6.5cm,height=6cm]{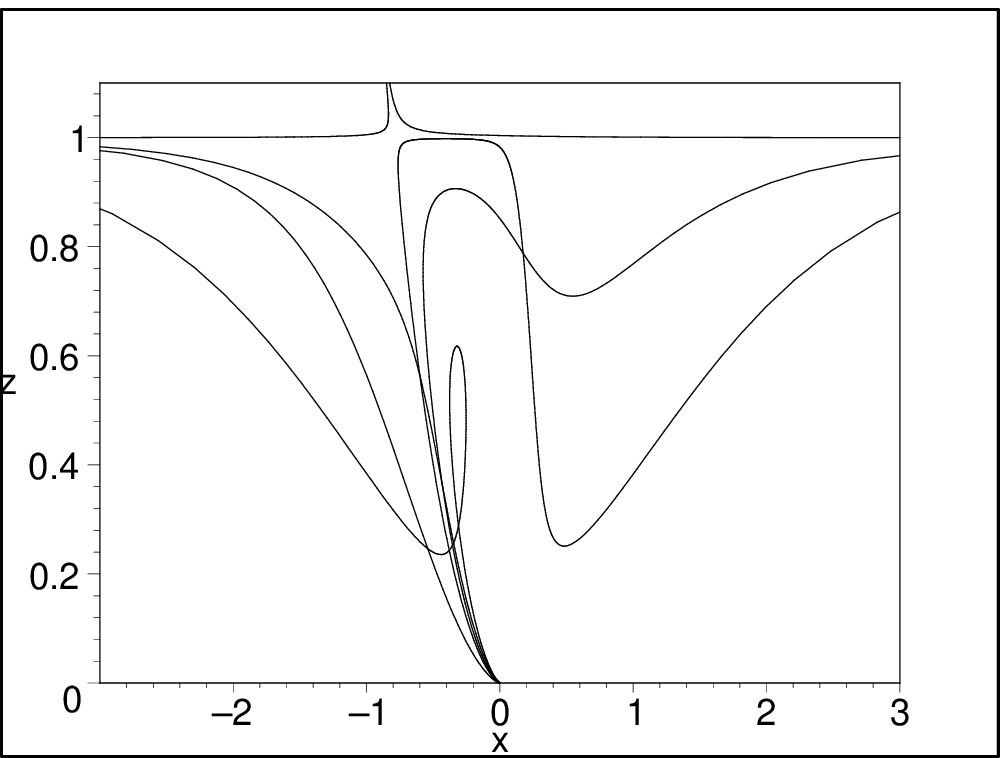}
\end{center}
\caption{Projections of the phase space trajectories for the scalar phantom field trapped in the RS brane, onto the phase planes $(x,z)$ and $(y,z)$, respectively. The brane tension $\lambda$ has been arbitrarily set equal to $\lambda=2$. Notice that not only the past asymptotic structure has been radically modified (there is now a past attractor: the empty Misner Universe), but also the future asymptotic has been modified: there is now no future attractor.} \label{fig2}
\end{figure}

\begin{figure}[t!]
\begin{center}
\includegraphics[width=6.5cm,height=6cm]{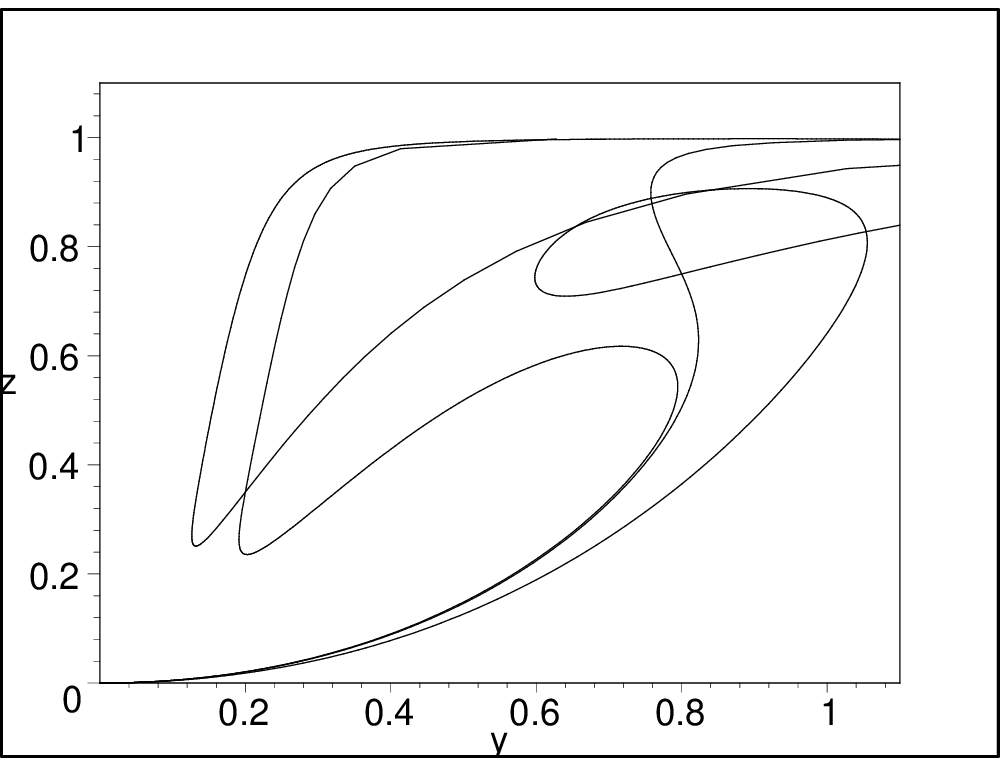}
\includegraphics[width=6.5cm,height=6cm]{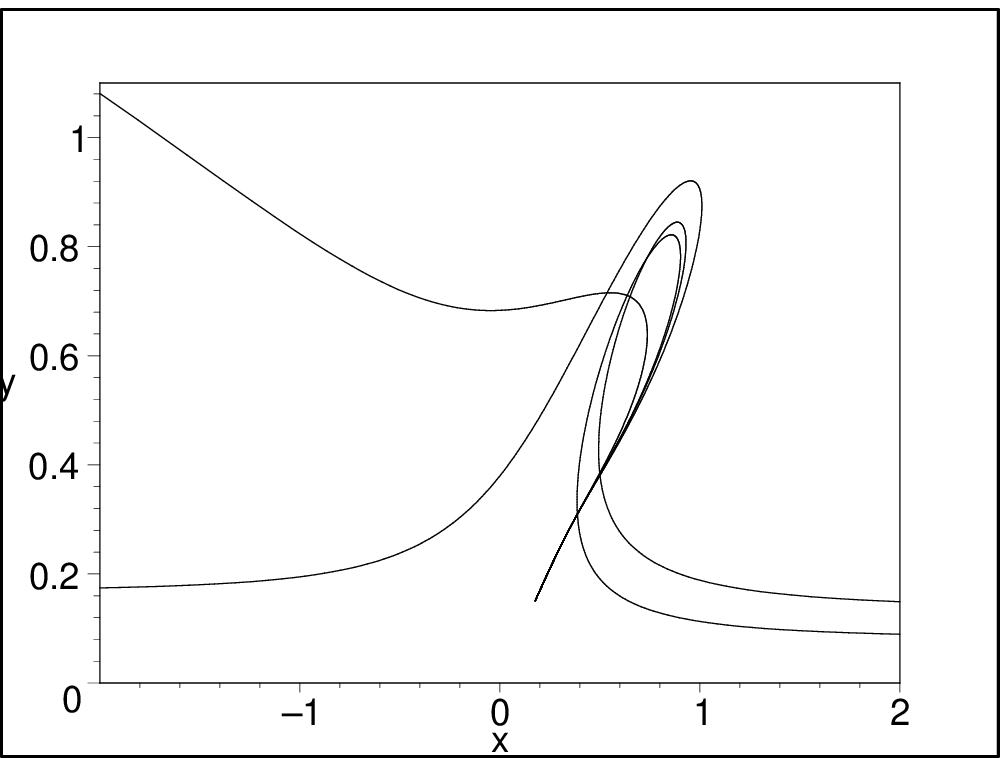}
\end{center}
\caption{Projections of the phase space trajectories for the tachyon phantom field trapped in the RS brane, onto the phase planes $(x,z)$ and $(y,z)$, respectively. The brane tension $\lambda$ has been arbitrarily set equal to $\lambda=2$. The empty Universe $(0,0,0)$ is the past attractor for any probe path in the phase space, while there is no any future attractor.} \label{fig3}
\end{figure}

\begin{table*}[tbp]\caption[crit]{Critical points of the autonomous systems of ODE (\ref{phantom-brane-ode}), and (\ref{T-brane-ode}), and their properties. These critical points always exist. The parameters $\lambda_1$, $\lambda_2$, and $\lambda_3$ are the eigenvalues of the linearization matrices corresponding to each one of the critical points. The upper half of the table is for the scalar phantom field, while the lower half is for the tachyon phantom field. We have defined: $y_*=\sqrt{(\lambda^2+\sqrt{\lambda^4+36})/6}$.}
\begin{tabular}{@{\hspace{14pt}}c@{\hspace{14pt}}c@{\hspace{14pt}}c@{\hspace{14pt}}c@{\hspace{14pt}}c@{\hspace{14pt}}c@{\hspace{14pt}}c@{\hspace{14pt}}c@{\hspace{14pt}}c@{\hspace{14pt}}c@{\hspace{14pt}}c}
\hline\hline\\[-0.3cm]
C. Point &$x$&$y$& $z$ &$\Omega_{cdm}$ &$\Omega_\phi$& $\omega_\phi$ & $q$ & $\lambda_1$&$\lambda_2$&$\lambda_3$\\[0.1cm]
\hline\\[-0.2cm]
$E$ & $0$ & $0$ & $0$ & $0$ & $0$ & undef. & $2$  &$6$& $3$ & $0$ \\[0.2cm]
$M$ & $0$ & $0$ & $1$ & $1$ & $0$ & undef. & $\frac{1}{2}$  &$-\frac{3}{2}$& $\frac{3}{2}$ & $-6$ \\[0.2cm]
$SP$&$-\frac{\lambda}{\sqrt 6}$&$\sqrt{1+\frac{\lambda^2}{6}}$ & $1$ &$0$ &$1$&$-1-\frac{\lambda^2}{3}$&$-1-\frac{\lambda^2}{2}$ &$-3-\lambda^2$&$-3-\frac{\lambda^2}{2}$ & $2\lambda^2$\\[0.2cm]
\hline \hline
$E$ & $0$ & $0$ & $0$ & $0$ & $0$ & $-1$ & $2$  &$3$& $3$ & - \\[0.2cm]
$M$ & $0$ & $0$ & $1$ & $1$ & $0$ & $-1$ & $\frac{1}{2}$  &$-3$&$\frac{3}{2}$ & $-3$ \\[0.2cm]
$TP$&$-\frac{\lambda y_*}{\sqrt 3}$ & $y_*$ & $1$ & $0$ & $1$ &$-1-\frac{\lambda^2y_*^2}{3}$&$-1-\frac{\lambda^2y_*^2}{2}$ &$-3-\frac{\lambda^2y_*^2}{2}$&$-3-\lambda^2y_*^2$ & $\lambda^2 y_*^2$ \\[0.2cm]
\hline \hline
\end{tabular}\label{tab2}
\end{table*}

\subsection{Brane effects}

Lets check now how the brane effects modify the dynamics of the scalar/tachyon phantom field. In this case we have to replace $\rho_B$ and $p_B$ in the cosmological equations (\ref{feqs''}) for matter trapped in the brane, by the corresponding expressions for the effective energy density and pressure for the scalar phantom field (\ref{phantom-rho-p}), and for the tachyon phantom field (\ref{T-rho-p}), respectively. In order to transform the corresponding field equations into an autonomous system of ODE one can use the same variables $x$, $y$ defined in (\ref{phantom-var}) for the scalar phantom field, or in (\ref{T-var}) for the tachyon phantom field, plus an additional variable:

\be z\equiv\frac{\rho_{tot}}{3H^2}.\label{z-var}\ee In terms of this variable $$\frac{\rho_{tot}}{\lambda}=\frac{2(1-z)}{z}\;\Rightarrow\;0<z\leq 1.$$ This means that general relativity is recovered at the value $z=1$. Notice that at the point $z=0$ the ratio $\rho_{tot}/\lambda$ is undefined, so that the plane set $\{(x,y,z)|\;z=0\}$ has to be removed from phase space. However, since we are interested in the asymptotic structure, we have to study, separately the limiting case $z\rightarrow 0$. 

For a scalar phantom field trapped in the RS brane the autonomous system of ODE that can be obtained out of the corresponding cosmological equations is the following (compare with (\ref{phantom-ode}), $V=V_0\exp{(-\lambda\phi)}$):

\bea &&x'=-\sqrt\frac{3}{2}\lambda y^2-3x-x\frac{H'}{H},\nonumber\\
&&y'=-\sqrt\frac{3}{2}\lambda xy-y\frac{H'}{H},\nonumber\\
&&z'=6(1-z)(z-x^2-y^2),\label{phantom-brane-ode}\eea where $$\frac{H'}{H}=-\frac{3}{2}\left(\frac{2-z}{z}\right)(z-x^2-y^2).$$ As before (scalar phantom field, no brane), for the phantom field dimensionless energy density parameter, and for the EOS parameter, one has $$\Omega_\phi=y^2-x^2,\;\;\omega_\phi=\frac{x^2+y^2}{x^2-y^2}.$$ However, the expressions for the CDM dimensionless energy density parameter, and for the deceleration parameter, differ from the ones that appear after equations (\ref{phantom-ode}): $$\Omega_{cdm}=z+x^2-y^2,\;\;q=-1+\frac{3}{2}\left(\frac{2-z}{z}\right)(z-x^2-y^2).$$ The phase space relevant for this case (recall that we consider only expanding Universes so that $y\geq 0$) is given by the following set: $$\Psi_{ph}^{brane}=\{(x,y,z)|\;y^2-x^2\geq 0,\;\;y\geq 0,\;\;0<z\leq 1\}.$$

In a similar fashion, for the tachyon phantom field on the RS brane we obtain the following autonomous system of ODE (compare with (\ref{T-ode}), $V=V_0 \phi^{-2}$):

\bea &&x'=-(x^2+1)(3x-\sqrt 3\lambda y),\nonumber\\
&&y'=-y\left(-\sqrt\frac{3}{4}\lambda xy+\frac{H'}{H}\right),\nonumber\\
&&z'=3(1-z)(z-y^2\sqrt{x^2+1}),\label{T-brane-ode}\eea where $$\frac{H'}{H}=-\frac{3}{2}\left(\frac{2-z}{z}\right)(z-y^2\sqrt{x^2+1}).$$ The cosmological parameters of observational interest are given by the following expressions: $$\Omega_\phi=\frac{y^2}{\sqrt{x^2+1}},\,\;\omega_\phi=-1-x^2,\;\;\Omega_{cdm}=z-\frac{y^2}{\sqrt{x^2+1}},$$ while, as customary, for the deceleration parameter: $q=-(1+H'/H)$. The phase space where to look for the equilibrium points of (\ref{T-brane-ode}) is $$\Psi_T^{brane}=\{(x,y,z)|\;z\sqrt{x^2+1}\geq y^2,\;\;y\geq 0,\;\;0<z\leq 1\}.$$

\subsubsection{Equilibrium points and their properties}

A simple inspection of equations (\ref{phantom-brane-ode}) and (\ref{T-brane-ode}), reveals that there can be a critical point lying on the phase plane $(x,y,0)$, which has been deliberately removed from the corresponding phase spaces. Actually, the point $(0,0,0)$ seems to be a critical point of each one of the autonomous systems described in both cases. Note, however, that at this point the expression for $H'/H$ is undefined, and so are first and second equations in (\ref{phantom-brane-ode}), and second equation in (\ref{T-brane-ode}). However this fact is a consequence of the chosen phase space variables. We can apply a secure approach to study the nature of the point $(0,0,0)$. In fact, we can expand the ODEs in (\ref{phantom-brane-ode},\ref{T-brane-ode}) in the neighborhood of $(0,0,0)$: $x\rightarrow 0+\epsilon_x$, $y\rightarrow 0+\epsilon_y$, $z\rightarrow 0+\epsilon_z$, where $(\epsilon_x,\epsilon_y,\epsilon_z)$ are the small perturbations. If we retain only the linear terms in the perturbations, i. e., if neglect terms $\epsilon_x^2\sim\epsilon_y^2\sim\epsilon_z^2\sim\epsilon_x\epsilon_y\sim\epsilon_x\epsilon_z\sim\epsilon_y\epsilon_z\sim 0$, and smaller, then the autonomous systems of ODE (\ref{phantom-brane-ode}), and (\ref{T-brane-ode}), transform into the following systems of linear ODE: $$\epsilon'_x=0,\;\;\epsilon'_y=3\epsilon_y,\;\;\epsilon'_z=6\epsilon_z,$$ and $$\epsilon'_x=\sqrt{3}\lambda\epsilon_y-3\epsilon_x,\;\;\epsilon'_y=3\epsilon_y,\;\;\epsilon'_z=3\epsilon_z,$$ respectively. Hence, in the case of the scalar phantom field trapped in the RS brane, the point $(0,0,0)$ is a non-hyperbolic critical point. Most we can say in this case is that it is unstable in the directions spanned by $\partial_y$ ($\epsilon_y\propto\exp{(3\tau)}$), and $\partial_z$ ($\epsilon_z\propto\exp{(6\tau)}$). However, the phase space drawings (Fig. \ref{fig2}) reveal that it is in fact a unstable critical point, which is the past attractor in $\Psi_{ph}^{brane}$. In the case of a tachyon phantom field living in the RS brane, the point $(0,0,0)$ is also the past attractor in the phase space $\Psi_T^{brane}$. In both cases this point is associated with $\Omega_{cdm}=\Omega_\phi=0$, so that it can be identified with an empty (Misner) Universe. The point $E=(0,0,0)$ has been included in Tab. \ref{tab2}.

Here we summarize the critical points of the autonomous systems of ODE (\ref{phantom-brane-ode}) and (\ref{T-brane-ode}), and their properties (these are also displayed in Tab. \ref{tab2}):

\begin{itemize}

\item{For the scalar phantom field trapped in the brane the critical points are:}

\begin{enumerate}
\item The empty Universe solution $E=(0,0,0)$, characterized by vanishing matter content $\Omega_{cdm}=\Omega_\phi=0$, is the past attractor. This is a super-decelerating solution with $q=2$. The effective EOS parameter of the phantom field $\omega_\phi$ is undefined.
\item The matter-dominated solution $M=(0,0,1)$ ($\Omega_{cdm}=1$, $\Omega_\phi=0$) is a saddle equilibrium point in $\Psi_{ph}^{brane}$. In this case the expansion is decelerated as well ($q=1/2$).
\item The phantom-dominated solution associated with the point $SP=(-\lambda/\sqrt{6},\sqrt{1+\lambda^2/6})$ ($\Omega_\phi=1$, $\omega_\phi<-1$), is also a saddle equilibrium point in the phase space $\Psi_{ph}^{brane}$ (the third eigenvalue is a positive quantity $\lambda_3=2\lambda^2$).
\end{enumerate}

\item{For the tachyon phantom field living in the RS brane the equilibrium points are the following:}

\begin{enumerate}
\item As in the previous case the empty (Misner) Universe $E=(0,0,0)$ ($\Omega_{cdm}=\Omega_\phi=0$), associated with super-decelerated pace of the expansion ($q=2$), is the past attractor in the phase space $\Psi_T^{brane}$.
\item The matter-dominated solution $M=(0,0,1)$ ($\Omega_{cdm}=1$, $\Omega_\phi=0$) is a saddle critical point, characterized by decelerating expansion ($q=1/2$).
\item The (tachyon) phantom-dominated solution $TP=(-\lambda y_*/\sqrt{3},y_*,1)$, where we have defined $y_*=\sqrt{(\lambda^2+\sqrt{\lambda^4+36})/6}$, is also a saddle critical point in $\Psi_T^{brane}$.
\end{enumerate}

\end{itemize}

\subsection{Concluding remarks of this section}

With the study of a standard (scalar) phantom model, and of a tachyon phantom scenario, we have corroborated what we expected. The extra-dimensional brane effects modify not only the past asymptotic structure, but also the future asymptotic properties of these phantom models. Actually, by comparing the structure of the phase space of the four-dimensional (standard) models in the first part of this section, with the one emerging after considering the brane effects (former subsection), one sees that (see the figures \ref{fig1}, \ref{fig2} and \ref{fig3}):

\begin{itemize}

\item While there were no past attractors in the phase spaces $\Psi_{ph}$, $\Psi_T$ corresponding to the models (\ref{phantom-ode}), (\ref{T-ode}), respectively, for the same matter content confined to a RS2 brane, there is a past attractor: the empty Misner Universe. Hence, the brane effects modify the past asymptotic (early-times) structure of the corresponding phase spaces.

\item The matter-dominated solution is not fundamentally affected by the brane effects.

\item While the phantom-dominated solution was the future attractor in the phase spaces $\Psi_{ph}$, $\Psi_T$ corresponding to the standard (scalar) phantom model, and to the tachyon phantom model, respectively, after brane effects are considered, the stability of the above solutions is modified: the phantom-dominated solution is now a saddle critical point, so that, in the corresponding brane models there is no future attractor in the phase space.

\end{itemize}

\begin{figure}[t!]
\begin{center}
\includegraphics[width=7.5cm,height=6cm]{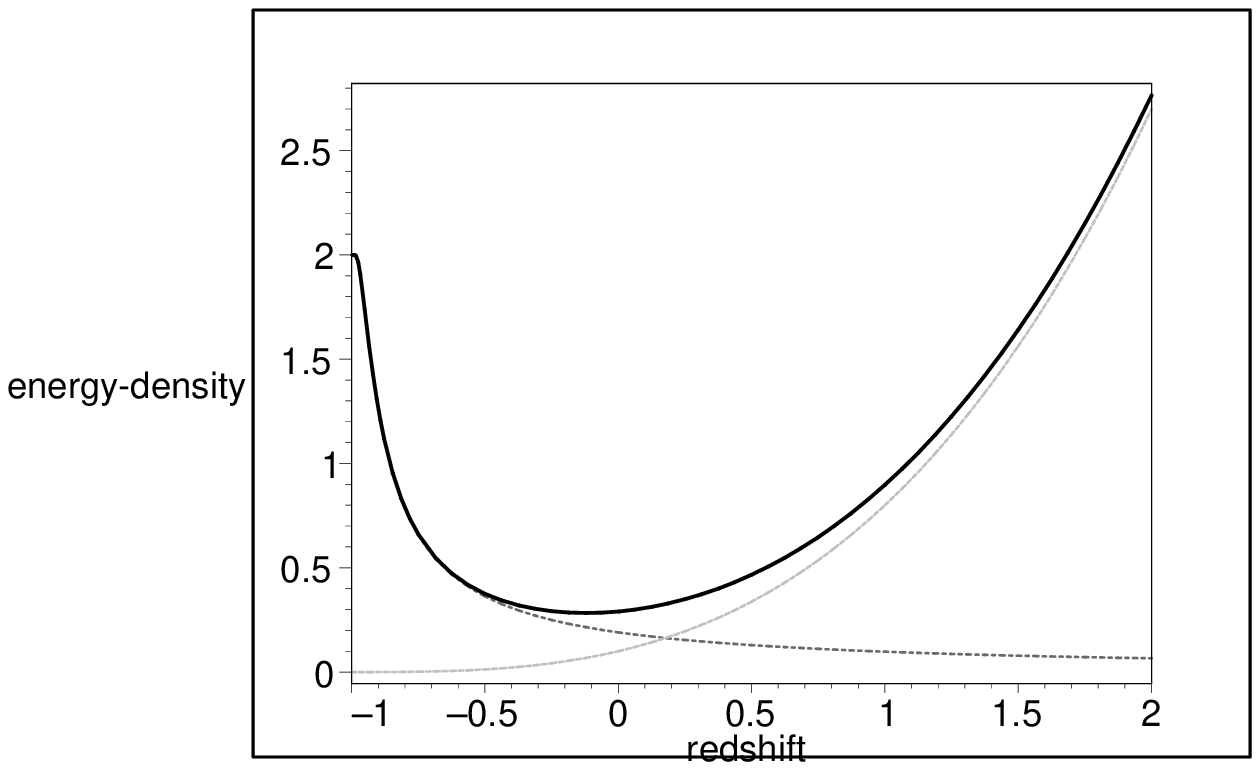}
\end{center}
\caption{Energy density vs redshift for the toy model $\rho_{tot}(z)=\mu(z+1)^3+\sigma(1-\exp{(-\nu(z+1)^{-1})})$ (Friedmann equation (\ref{toy model})). The free parameters have been arbitrarily chosen: $\mu=0.1$, $\sigma=2$, $\nu=0.1$. The solid curve represents the evolution of the total energy density, while the darker dashed line represents the phantom energy density (the cold dark matter energy density evolution is represented by the gray dashed line).} \label{fig4}
\end{figure}

\section{Discussion and Conclusion}

Conventional wisdom convinces us that Randall-Sundrum brane scenario, which was proposed as an alternative to the Kaluza-Klein compactification procedure -- also to seek for an alternative explanation to the mass hierarchy problem --, appreciably modifies gravity only at very high energy/short scales (UV modifications only), so that it could have impact on the pace of the primordial inflation, for instance, but does not modify the late-time cosmic dynamics. However, the above statement is true only if the energy density of the matter content of the brane, dilutes with the course of the cosmic expansion, as it is true for most forms of known matter sources (radiation, dust, quintessential dark energy, etc.). What happens if those known forms of matter, are replaced by other (exotic) forms which energy density does not dilute, but increases at late times? Take, for instance, a Universe filled with a mixture of cold dark matter and of a dark energy fluid which energy density grows with the cosmic expansion according to $\rho_{ph}=\sigma(1-e^{-\nu a})$, so that the total energy density of the matter trapped in the RS brane is given by:

\be \rho_{tot}=\frac{\mu}{a^3}+\sigma(1-e^{-\nu a}).\label{toy model}\ee The behavior of the total energy density $\rho_{tot}$ vs the redshift $z$ is shown in the figure \ref{fig4}. It is seen that the cold dark matter dominates at early times and dilutes as the expansion proceeds. Meanwhile, the phantom (dark energy) density is a growing function of the redshift and, somewhere in the future starts dominating the dynamics of the expansion. The fate of the cosmic evolution ($z\rightarrow -1$) is close to a de Sitter Universe where $\rho_{tot}\approx\sigma$. Although this is an unrealistic toy model, several qualitative aspects will be well illustrated. Since the brane effects modify the Friedmann equation according to $$3H^2=\rho_{tot}\left(1+\frac{\rho_{tot}}{2\lambda}\right),$$ then, at very early times where $z\gg (2\lambda/\mu)^{1/3}$, the Friedmann equation is well approximated by $3H^2=\mu^2/2\lambda a^6$. This means that the effect of the brane at high energies and large redshifts, is to allow for a metamorphosis of the background fluid from cold dark matter into a super-relativistic stiff fluid.  At late times the modification to the standard Eisntein's dynamics is also appreciable. Actually, at late times, according to standard Friedmann equation $3H^2=\rho_{tot}\approx\sigma$, meanwhile, if brane effects are taken into consideration, then, either $3H^2\approx 3\sigma/2$ if $\sigma$ is of the same order as the brane tension ($\sigma\approx\lambda$), or $3H^2\approx\sigma^2/2\lambda$ if $\sigma\gg\lambda$. In the former case, aside from the factor $3/2$, it has to be considered, also, that $\sigma\approx\lambda$ is a very large quantity. In the latter case the energy density is boosted by the factor $\sigma/2\lambda$ which is also very large since $\sigma/\lambda\gg 1$. This shows the way brane effects might affect the cosmic dynamics not only at early times but also at late times.

While the toy model (\ref{toy model}) is a unrealistic model with no physical motivation at all, other more physically motivated models, as the scalar phantom, and tachyon phantom models, also corroborate the former results. Even in the case when the effective phantom behavior is the result of the peculiar dynamics of a cosmological fluid of known nature (the NLED-based model of section II), the extra-dimensional brane effects modify the late time dynamics. In all of these models of very different nature, the effect of the brane is to modify the stability of the solution which is associated with the future asymptotics of the models: the brane destroys the future attractor, leaving in its place a unstable saddle equilibrium point. This means that the end point of the cosmic evolution is uncertain, unlike the case when there is a future attractor which is, necessarily, the end point of the cosmological evolution. It has to be underlined that the modification of the late-time asymptotic structure by the RS brane effects occurs only for those exotic forms of matter which effective equation of state in GR is less than -1 (phantom matter).

We want to stress that the above "mixing of scales" effect, where both UV and IR modifications of general relativity are the result of a same effect (in the present case the effect of the extra-space in the Randall-Sundrum brane picture), is distinctive only of theories that modify the right-hand-side (matter part) of the Friedmann equation. In this sense, for instance, it is not to be expected that the DGP brane model -- designed to modify gravity at the IR -- can modify GR also at UV energies/curvatures. This will be the subject of a separate publication \cite{extenso}.

The authors greatly acknowledge most useful comments and criticism by the referees. This work was partly supported by CONACyT M\'{e}xico, under grants 49924-J, 105079, and Instituto Avanzado de Cosmologia (IAC) collaboration. R. G.-S. acknowledges partial support from COFAA-IPN and EDI-IPN grants, and SIP-IPN 20100610, 20110664.

\end{document}